\documentclass{article}
\usepackage{times}
\begin{document}
\title{Quantum--Mechanical Dualities\\ from Classical Phase Space}
\author{Jos\'e M. Isidro\\
Instituto de F\'{\i}sica Corpuscular (CSIC--UVEG)\\
Apartado de Correos 22085, Valencia 46071, Spain\\
{\tt jmisidro@ific.uv.es}}
\hyphenation{Mi-nis-te-rio}

\maketitle

\begin{abstract}

The geometry of the classical phase space ${\cal C}$ of a finite number of degrees of 
freedom determines the possible duality symmetries of the corresponding 
quantum mechanics. Under {\it duality}\/ we understand the relativity of the 
notion of a quantum with respect to an observer on ${\cal C}$. 
We illustrate this property explicitly in the case when 
classical phase space is complex $n$--dimensional projective space ${\bf CP}^n$. 
We also provide some examples of classical  dynamics on ${\bf CP}^n$
that exhibit these properties at the quantum level.


2001 PACS codes: 03.65.Bz, 03.65.-w, 03.65.Ca, 03.65.Sq.

\end{abstract}

\tableofcontents

\section{Introduction}\label{intro}

Duality can be understood as the relativity of the notion of a quantum \cite{VAFA}.
Following ref. \cite{VAFA}, a framework is needed that can implement dualities 
in the quantum mechanics of a finite number of degrees of freedom \cite{HOLLAND}. 
The purpose of this letter is to expand on previous work \cite{PLA, PQM},
where such a framework has been presented. We would also like to draw attention 
to refs. \cite{MATONE, DAVIS, ANANDAN}.

The notion of duality implies the possibility of nontrivial transformations between
the Hilbert spaces of quantum states corresponding to different observers on classical 
phase space ${\cal C}$. Under an {\it observer}\/ one understands, in general 
relativity, a little man carrying a ruler and a clock. In fact one may 
forget about the little man while keeping his ruler and his clock, to conclude that 
an observer is just a local coordinate chart on spacetime. 
(We adopt the active point of view, under which a coordinate change
corresponds to a transformation between different spacetime points).
{\it Mutatis mutandi}, in a quantum--mechanical setup, an observer will be a local  
coordinate chart $({\cal U}, z)$ on ${\cal C}$, where ${\cal U}$ 
is an open neighbourhood of a point $p\in{\cal C}$ endowed with the coordinate functions $z$.
We intend to develop the quantum theory corresponding to this observer. 
Pick a vacuum state $|0(p)\rangle$ (obtained by minimising a certain 
Hamiltonian on ${\cal U}$). Under a coordinate change from 
$p$ to $\tilde p$ we have $({\cal U}, z)\rightarrow (\tilde {\cal U}, \tilde z)$, 
and the vacuum changes as $|0(p)\rangle\rightarrow|0(\tilde p)\rangle$. Performing this operation 
sufficiently many times we succeed in covering ${\cal C}$ with a family of 
1--dimensional vector spaces, pointwise generated by the corresponding 
vacua. Geometrically we have defined a complex line bundle $N({\cal C})$ over ${\cal C}$, 
whose fibre over $p\in {\cal C}$ is generated by $|0(p)\rangle$.

Corresponding to $N({\cal C})$ we can erect a Hilbert--space fibre ${\cal H}(p)$
over $p\in {\cal C}$ by considering all possible quantum excitations of $|0(p)\rangle$.
Letting then $p$ vary over ${\cal C}$,  we define a complex vector bundle ${\cal QH}({\cal C})$
that we call {\it quantum Hilbert--space bundle}, or just ${\cal QH}$--bundle for short.

Now it is clear that different observers on ${\cal C}$ may be quantised 
differently, even after a fixed choice for the vacuum line bundle $N({\cal C})$.
If ${\cal QH}({\cal C})$ is a flat bundle, parallel transport allows to 
canonically identify the fibres  ${\cal H}(p)$,  ${\cal H}(\tilde p)$ over different points.
(The connection itself will be univocally defined by requiring it to be metric--compatible 
and torsion--free). However, on a nonflat bundle there is no such canonical identification 
between different fibres; there is also no {\it a priori}\/ reason for ${\cal QH}({\cal C})$
to be flat. Only on certain manifolds ${\cal C}$ can we ensure that ${\cal QH}({\cal C})$ 
will be flat ({\it e.g.}, when ${\cal C}$ is contractible). 
In the general case ${\cal QH}({\cal C})$ will be nonflat. 
The absence of a canonical identification between the fibres above different points allows 
for duality transformations. For example, one such imaginable duality could be having a semiclassical 
state $\vert\psi(p)\rangle\in {\cal H}(p)$ transform into a highly quantum excitation 
$\vert\psi(\tilde p)\rangle\in{\cal H}(\tilde p)$ when passing from $p$ to $\tilde p$.
The precise nature of the specific dualities so obtained will depend on the 
geometry of ${\cal C}$ and ${\cal QH}({\cal C})$.

Another possible implication of duality is the following. Quantum excitations 
may be measured with respect to different vacua. 
The corresponding quanta are physically inequivalent. Allowing for more than 
one, physically inequivalent, vacuum is possible if ${\cal C}$ admits
more than one equivalence class of complex line bundles $N({\cal C})$.

Our notations are as follows (see refs. \cite{LIBSCHLICHENMAIER, ARNOLD, LIBAZCA} for 
background material). ${\cal C}$ will denote a complex $n$--dimensional,
compact classical phase space, endowed with a symplectic form $\omega$ 
and a complex structure ${\cal J}$. We will assume that $\omega$ and ${\cal J}$ 
are compatible, so holomorphic coordinate charts on ${\cal C}$ are also 
Darboux charts. On a complex manifold ${\cal C}$, the Picard group ${\rm 
Pic}\, ({\cal C})$ classifies holomorphic equivalence classes of complex 
line bundles $N({\cal C})$.

We will concentrate on the case when ${\cal C}$ is complex projective space ${\bf CP}^n$. 
Then we have ${\rm Pic}\, ({\bf CP}^n)={\bf Z}$. In this way inequivalent, 
holomorphic line bundles $N_l({\bf CP}^n)$ over ${\bf CP}^n$ are 1--to--1 with the integers 
$l\in{\bf Z}$. The following holomorphic vector bundles 
over ${\bf CP}^n$ will be considered: the hyperplane bundle $\tau$ and its dual $\tau^{*}$, 
also called the tautological line bundle; the holomorphic tangent bundle $T({\bf CP}^n)$ 
and its dual $T^*({\bf CP}^n)$. Every holomorphic line bundle $N_l({\bf CP}^n)$ over ${\bf CP}^n$ 
is isomorphic to $\tau ^l$ for some $l\in {\bf Z}$; this integer is the Picard class.  
The fibrewise generator of $N_l({\bf CP}^n)$ is the vacuum state 
$\vert 0\rangle_l$. Compactness of ${\bf CP}^n$ implies that the Hilbert space ${\cal H}$ 
is finite--dimensional; we will show that the minimum value of ${\rm dim}\,{\cal H}$ is $n+1$.

\section{Quantum Hilbert--space bundles over ${\bf CP}^n$}\label{esstqmb}

\subsection{Construction of the bundle ${\cal QH}_{l=1}({\bf CP}^n)$}\label{xcompt}

Let $Z^1,\ldots, Z^{n+1}$ denote homogeneous coordinates on ${\bf CP}^n$. 
The chart defined by $Z^k\neq 0$ covers one copy of the open set 
${\cal U}_k={\bf C}^n$. On the latter we have the holomorphic coordinates 
$z^j_{(k)}=Z^j/Z^k$, $j\neq k$; there are $n+1$ such coordinate charts. 

To begin with we will choose the Picard class $l=1$.
Starting from ${\cal C}={\bf CP}^{0}$, {\it i.e.}, a point $p$ as classical phase space, 
the space of quantum rays must also reduce to a point. Then the corresponding Hilbert space 
is ${\bf C}$. The only state is the vacuum $|0\rangle_{l=1}$.

Next we pass from ${\cal C}={\bf CP}^0$ to ${\cal C}={\bf CP}^1$. 
Regard $p$, henceforth denoted $p_1$, as the {\it point at infinity}\/ 
with respect to a coordinate chart $({\cal U}_1, z_{(1)})$ on ${\bf CP}^1$ that does not 
contain $p_1$. This chart is biholomorphic to ${\bf C}$ and supports a representation 
of the Heisenberg algebra in terms of creation and annihilation operators $A^{\dagger}(1)$,  
$A(1)$. This process adds the new state $A^{\dagger}(1)|0\rangle_{l=1}$ to the spectrum.
The new Hilbert space ${\bf C}^2$ is the linear span of $|0\rangle_{l=1}$ 
and $A^{\dagger}(1)|0\rangle_{l=1}$.

On ${\bf CP}^1$ we have the charts $({\cal U}_1, z_{(1)})$ and $({\cal U}_2, z_{(2)})$. 
Point $p_1$ is at infinity with respect to $({\cal U}_1, z_{(1)})$, while it 
belongs to $({\cal U}_2, z_{(2)})$. Similarly, the point at infinity with respect to 
$({\cal U}_2, z_{(2)})$, call it $p_2$, belongs to $({\cal U}_1, z_{(1)})$ but not to 
$({\cal U}_2, z_{(2)})$. Above we have proved that the ${\cal QH}$--bundle
has a fibre ${\bf C}^2$ which, on the chart ${\cal U}_1$, is the linear span 
of $|0\rangle_{l=1}$ and $A^{\dagger}(1)|0\rangle_{l=1}$. On the chart ${\cal U}_2$, the 
fibre is the linear span of $|0\rangle_{l=1}$ and $A^{\dagger}(2)|0\rangle_{l=1}$, 
$A^{\dagger}(2)$ being the creation operator on ${\cal U}_2$. On the common overlap 
${\cal U}_1\cap {\cal U}_2$, the coordinate transformation between $z_{(1)}$ and 
$z_{(2)}$ is holomorphic. This implies that, on ${\cal U}_1\cap {\cal U}_2$,
the fibre ${\bf C}^2$ can be taken in either of two equivalent ways: either 
as the linear span of $|0\rangle_{l=1}$ and $A^{\dagger}(1)|0\rangle_{l=1}$, or as that 
of $|0\rangle_{l=1}$ and $A^{\dagger}(2)|0\rangle_{l=1}$. 

The general construction is now clear. Topologically we have ${\bf CP}^{n}={\bf 
C}^n\cup {\bf CP}^{n-1}$, with ${\bf CP}^{n-1}$ a hyperplane at infinity, 
but we also need to describe the coordinate charts and their overlaps.
There are coordinate charts $({\cal U}_k, z_{(k)})$, $k=1, \ldots, n+1$ 
and nonempty $f$--fold overlaps $\cap_{k=1}^f {\cal U}_k$ for $f=2,3,\ldots, n+1$. 
Each chart $({\cal U}_k, z_{(k)})$ is biholomorphic with ${\bf C}^n$ and has
a ${\bf CP}^{n-1}$--hyperplane at infinity; the latter is charted by the 
remaining charts $({\cal U}_r, z_{(r)})$, $r\neq k$. Over $({\cal U}_k, z_{(k)})$ 
the Hilbert--space bundle ${\cal QH}$ has a fibre ${\bf C}^{n+1}$ spanned by 
\begin{equation}
|0\rangle_{l=1},\qquad  A_i^{\dagger}(k) |0\rangle_{l=1}, \qquad i=1,2,\ldots, n. 
\label{pann}
\end{equation}
Analyticity arguments similar to those above prove that, on every nonempty 
$f$--fold overlap $\cap_{k=1}^f {\cal U}_k$, the fibre ${\bf C}^{n+1}$ can 
be taken in $f$ different, but equivalent ways, as the linear span of 
$|0\rangle_{l=1}$ and $A_i^{\dagger}(k) |0\rangle_{l=1}$,  $i=1,2,\ldots, n$, for 
every choice of $k=1,\ldots, f$.

For the transition functions we take the following. The vacuum state $|0\rangle_{l=1}$
transforms with the transition function $t(\tau)$ corresponding to 
the line bundle $N_{l=1}({\bf CP}^n)=\tau$.  
The excited states $A_i^{\dagger}(k) |0\rangle_{l=1}$ will transform according to $n\times n$ 
jacobian matrices $j({\bf CP}^n)$ for the coordinate changes on ${\bf CP}^n$. 
In this way the $A_i^{\dagger}(k) |0\rangle_{l=1}$ provide (fibrewise)
a basis for the tangent bundle $T({\bf CP}^n)$.
Altogether, the transition functions $t\left({\cal QH}_{l=1}({\bf 
CP}^n)\right)$ decompose as the direct sum
\begin{equation}
t\left({\cal QH}_{l=1}({\bf CP}^n)\right)=j({\bf CP}^n)\oplus t(\tau),
\label{labastidahijoputa}
\end{equation}
and the complete bundle is
\begin{equation}
{\cal QH}_{l=1}({\bf CP}^n)=T({\bf CP}^n)\oplus \tau.
\label{jodetecabronlabastida}
\end{equation}
We have so far assumed that $l=1$. The case $l=0$ corresponds to the 
trivial line bundle while, for $l=-1$, the previous construction 
holds throughout if we replace every bundle with its dual. 
Thus, on the chart ${\cal U}_k$, $k=1,\ldots, n+1$, the dual fibre 
is the linear span of 
\begin{equation}
_{l=1}\langle 0|,\qquad  _{l=1}\langle 0|A_i(k), \qquad i=1,2,\ldots, n.
\label{labastidamaricondeplaya}
\end{equation}
The case $\vert l\vert>1$ will be treated in section \ref{labastidamecagoentuputamadre}.

\subsection{Diagonalisation of the projective Hamiltonian}\label{tnclmm}

Deleting from ${\bf CP}^n$ the ${\bf CP}^{n-1}$--hyperplane at infinity produces the noncompact 
space ${\bf C}^n$, which is the classical phase space of the $n$--dimensional harmonic 
oscillator (now no longer {\it projective}\/, but {\it linear}\/). The corresponding Hilbert space 
${\cal H}$ is infinite--dimensional because the symplectic volume of ${\bf C}^n$ is infinite.

The deletion of the hyperplane at infinity may be understood from the 
viewpoint of the K\"ahler potential corresponding to the 
Fubini--Study metric on ${\bf CP}^n$.
On the chart $({\cal U}_k, z_{(k)})$ the K\"ahler 
potential reads
\begin{equation}
K(z^j_{(k)}, {\bar z}^j_{(k)})=
\log{\left(1 + \sum_{j=1}^n z^j_{(k)} {\bar z}^j_{(k)}\right)}.
\label{fubst}
\end{equation} 
No longer being able to pass holomorphically 
from a point at finite distance to a point at infinity implies that, 
on the conjugate chart $({\cal U}_k, z_{(k)})$, the squared modulus 
$|z_{(k)}|^2$ is always small and we can Taylor--expand eqn. (\ref{fubst}) as
\begin{equation}
\log{\left(1 + \sum_{j=1}^n z^j_{(k)} {\bar z}^j_{(k)}\right)}\simeq 
\sum_{j=1}^n z^j_{(k)} {\bar z}^j_{(k)}.
\label{tayy}
\end{equation}
The right--hand side of eqn. (\ref{tayy}) is the K\"ahler potential for 
the usual Hermitean metric on ${\bf C}^n$. As such, $\sum_{j=1}^n z^j_{(k)} {\bar 
z}^j_{(k)}$ equals the classical Hamiltonian for the $n$--dimensional linear harmonic 
oscillator. Observers on this coordinate chart effectively see ${\bf C}^n$ 
as their classical phase space. The corresponding Hilbert space is the 
(closure of the) linear span of the states $|m_1,\ldots, m_n\rangle$, where 
\begin{equation}
H_{\rm lin}|m_1,\ldots, m_n\rangle = \sum_{j=1}^n 
\left(m_j + {1\over 2}\right)|m_1,\ldots, m_n\rangle,\qquad 
m_j=0,1,2,\ldots,
\label{oegg}
\end{equation}
and 
\begin{equation}
H_{\rm lin}=\sum_{j=1}^n\left(A^{\dagger}_j(k)A_j(k)+{1\over 2}\right)
\label{hachelin}
\end{equation} 
is the quantum Hamiltonian operator corresponding to the classical 
Hamiltonian function on the right--hand side of eqn. (\ref{tayy}). Then the stationary 
Schr\"odinger equation for the {\it projective}\/ oscillator reads
\begin{equation}
H_{\rm proj}|m_1,\ldots, m_n\rangle = \log\left(1+\sum_{j=1}^n 
\left(m_j + {1\over 2}\right)\right)|m_1,\ldots, m_n\rangle,
\label{oeggpp}
\end{equation}
where 
\begin{equation}
H_{\rm proj}=\log\left(1+\sum_{j=1}^n\left(A^{\dagger}_j(k)A_j(k)+
{1\over 2}\right)\right)
\label{hacheproj}
\end{equation}
is the quantum Hamiltonian operator corresponding to 
the classical Hamiltonian function on the left--hand side of eqn. (\ref{tayy}).

The same states $|m_1,\ldots, m_n\rangle$ that diagonalise $H_{\rm lin}$ also 
diagonalise $H_{\rm proj}$. However, eqns. (\ref{oegg})--(\ref{hacheproj}) 
above in fact only hold locally on the chart ${\cal U}_k$, which does not cover 
all of ${\bf CP}^n$. Bearing in mind that there is one hyperplane at infinity 
with respect to this chart, we conclude that the arguments of section \ref{xcompt}
apply in order to ensure that the projective oscillator only has $n$ excited states.
Then the occupation numbers $m_j$ are either all 0 (for the vacuum state) 
\begin{equation}
\vert 0\rangle_{l=1}=\vert m_1=0, \ldots, m_n=0\rangle,
\label{mariconzonlabastida}
\end{equation}
or all zero but for one of them, where $m_i=1$ (for the excited states)
\begin{equation}
A_i(j)^{\dagger}\vert 0\rangle_{l=1}=\vert m_1=0, \ldots, m_i=1, \ldots, m_n=0\rangle, 
\qquad i=1, \ldots, n,
\label{labastidacomemierda}
\end{equation}
and ${\rm dim}\,{\cal H}=n+1$ as it should.

One further conclusion that we can draw from the Hamiltonian analysis is 
that the vacuum $\vert 0\rangle_{l=1}$ of $H_{\rm proj}$ is nondegenerate. 
This is so because the 
$H_{\rm lin}$ in eqn. (\ref{hachelin}) has a 
nondegenerate vacuum, and the logarithm in eqn. (\ref{hacheproj}) 
is a monotonically increasing function. Therefore the parameter space 
for physically inequivalent vacua is correctly given by the Picard group. Indeed the 
latter classifies inequivalent {\it line}\/ bundles. This conclusion 
might appear unnecessary, since we know from the textbooks
that the ground state is nondegenerate. However the nondegeneracy 
of the vacuum of $H_{\rm proj}$ on ${\bf CP}^n$ was by no means 
guaranteed, as the standard proof of nondegeneracy of the vacuum 
goes back to a mathematical theorem applicable to second--order differential 
operators \cite{CH}. Our Hamiltonian $H_{\rm proj}$ is not second order.

\subsection{Construction of the bundle ${\cal QH}_{(\rho(l), l)}({\bf CP}^n)$}
\label{labastidamecagoentuputamadre}

In section \ref{xcompt} we have constructed a bundle of $(n+1)$--dimensional Hilbert spaces 
for the Picard class $l=1$.  Now eqns. (\ref{labastidahijoputa}), (\ref{jodetecabronlabastida})
imply that this vector bundle has $SU(n)\times U(1)$ as its structure group, of which 
the $(n+1)$--dimensional representation eqn. (\ref{pann}) provides the defining representation
(and eqn. (\ref{labastidamaricondeplaya}) its dual).

Any representation $\rho$ of $SU(n+1)\supset SU(n)\times U(1)$ restricts to 
a representation of $SU(n)$, that we continue to denote by $\rho$.
We could pick any such representation, 
plus a Picard class $l\in {\bf Z}$, to construct a ${\cal QH}$--bundle 
\begin{equation}
{\cal QH}_{(\rho, l)}({\bf CP}^n)=\rho(T{\bf CP}^n)\oplus 
\tau^l.
\label{ramallocasposo}
\end{equation}
Under coordinate changes on ${\bf CP}^n$, the vacuum $\vert 0\rangle_l$ transforms 
with $t^l(\tau)$, while its excitations transform according to jacobian matrices
$j_{\rho}({\bf CP}^n)$ expressed in the representation $\rho$. 
Corresponding to eqn. (\ref{ramallocasposo}) we would have the transition 
functions
\begin{equation}
t\left({\cal QH}_{(\rho, l)}({\bf CP}^n)\right)=j_{\rho}({\bf CP}^n)\oplus t^l(\tau).
\label{labastidahijoperra}
\end{equation}
In principle, eqns. (\ref{ramallocasposo}) and (\ref{labastidahijoperra}) 
would provide the most general ${\cal QH}$--bundles that one can consider on ${\bf CP}^n$, 
were it not for the following reason. In our framework, tangent vectors are
quantum states obtained as excitations of the vacuum. As such, a tangent 
vector in representation $\rho$ cannot be completely arbitrary.
It must reflect the fact that it is the result of acting on the vacuum $\vert 0\rangle_l$ 
with a creation operator $A^{\dagger}$. This requirement imposes some 
constraints on the representation: $\rho$ must be a 
function of the Picard class $l$ determining the vacuum $\vert 
0\rangle_l$. The precise dependence $\rho=\rho(l)$ has been 
obtained in ref. \cite{PQM}, and will be summarised next following an 
alternative, though equivalent, argument.

Replacing ${\bf CP}^n$ in section \ref{tnclmm} with ${\bf CP}^{n+l}$, 
where $l>1$ is the Picard class under consideration on ${\bf CP}^n$, the vacuum 
corresponding to the Picard class $l'=1$ on ${\bf CP}^{n+l}$ has $n+l$ 
zero occupation numbers, 
\begin{equation}
\vert 0\rangle_{l'=1}=\vert m_1=0, \ldots, m_{n+l}=0\rangle.
\label{ramallomariconzon}
\end{equation}
On ${\bf CP}^{n+l}$ there are $n+l$ creation operators $A_i^{\dagger}$, 
$i=1, \ldots, n+l$. Picking $n$ out of these and acting with them on 
$\vert 0\rangle_{l'=1}$ we obtain one state that, as seen from ${\bf 
CP}^{n+l}$, is excited. However, as seen from ${\bf CP}^{n}$, 
that same state is either the vacuum $\vert 0\rangle_{l}$ or one of its excitations. 
There are $\left(n+l\atop n\right)$ such independent states. 
Such is the dimension of the space of states on ${\bf CP}^{n}$ 
corresponding to the Picard class $l>1$.

$SU(n+l)$ Young tableaux with a single column of $n$ boxes correspond to
representations of $SU(n+l)$ with dimension $\left(n+l\atop n\right)$.
By restriction they are also representations of $SU(n+1)$ and, ultimately, 
of $SU(n)$. Such are the allowed representations $\rho$; as advanced 
earlier, they depend on the Picard class $l$. It should be borne in mind that 
the states in these representations are {\it not}\/ of the form $A_i^{\dagger}\vert 
0\rangle_l$, with $A_i^{\dagger}$, $i=1, \ldots, n$, a creation operator on ${\bf CP}^{n}$.
Rather, they are of the form $A_i^{\dagger}\vert 0\rangle_{l'=1}$, 
with $A_i^{\dagger}$, $i=1, \ldots, n+l$, a creation operator on ${\bf CP}^{n+l}$.
Thus the vacuum $\vert 0\rangle_{l}$, $l>1$, is actually {\it degenerate}\/ 
on ${\bf CP}^n$, since it arises as an {\it excited}\/ state with respect 
to the vacuum $\vert 0\rangle_{l'=1}$ on ${\bf CP}^{n+l}$. However, even 
if $\vert 0\rangle_{l}$, $l>1$, is a degenerate vacuum, it continues to 
hold that ${\rm Pic}\, ({\bf CP}^n)={\bf Z}$ is the parameter space for 
physically inequivalent vacua.

As a consistency check, setting $l=1$ reproduces the dimension $n+1$ of the defining representation. 
With these conditions on $\rho$, eqns. (\ref{ramallocasposo}) and (\ref{labastidahijoperra}) 
are finally correct.

\section{Discussion}\label{dsskk}

We have argued that the geometry of classical phase space ${\cal C}$ determines the 
possible duality symmetries of the corresponding quantum mechanics.
To this end we have introduced the Picard group ${\rm Pic}({\cal C})$ as the parameter space for 
physically inequivalent vacua. This allows to construct different, 
inequivalent families of Hilbert--space bundles of quantum states over ${\cal C}$. 
In particular, quantum states other than the vacuum appear as tangent 
vectors to ${\cal C}$.

Our analysis has dealt with the case when ${\cal C}={\bf CP}^n$, for which
we have given a classification of all possible ${\cal QH}$--bundles.
A mathematical pendulum with an $(n+1)$--dimensional configuration space
has a {\it reduced}\/ phase space given by ${\bf CP}^n$, where the 
reduction consists in using the energy (a constant of the motion)
to decrease the number of degrees of freedom by one \cite{ARNOLD}. We have also 
given in section \ref{tnclmm} another example of a classical dynamics on ${\bf CP}^n$.

As a complex manifold, ${\bf CP}^n$ admits a Hermitian metric, so having tangent vectors 
as quantum states suggests using the Hermitian connection and the corresponding 
curvature tensor to measure flatness. The freedom in having different nonflat Hilbert--space
bundles over ${\bf CP}^n$ resides in the different possible choices for the complex line bundle 
$N_l({\bf CP}^n)$.  Every choice of a vacuum leads to a different set of excitations 
and thus to a different quantum mechanics. Moreover, the ${\cal QH}$--bundles constructed 
here are nonflat. This implies that, even after fixing a vacuum,
there is still room for duality transformations between different observers 
on classical phase space. These two facts provide an explicit 
implementation of quantum--mechanical dualities.

{\bf Acknowledgements}

It is a great pleasure to thank J. A. de Azc\'arraga for encouragement and support.
This work has been partially supported by research grant BFM2002--03681 from 
Ministerio de Ciencia y Tecnolog\'{\i}a and EU FEDER funds.


\begin{thebibliography}{99}


\bibitem{VAFA}
C. Vafa, {\tt hep-th/9702201}.

\bibitem{HOLLAND}
P. Holland, {\it The Quantum Theory of Motion}, Cambridge University 
Press, Cambridge (1993).

\bibitem{PLA}
J.M. Isidro, {\it Phys. Lett.} {\bf A301} (2002) 210.

\bibitem{PQM}
J.M. Isidro, {\tt hep-th/0304175}; {\tt hep-th/0304235}.

\bibitem{MATONE}  
A. Faraggi and M. Matone, {\it Phys. Lett.} {\bf A249} (1998) 180;\\
G. Bertoldi, A. Faraggi and M. Matone, {\it Class. Quant. Grav.} {\bf 17} (2000) 3965.

\bibitem{DAVIS}
E. Davis and G. Ghandour, {\it Phys. Lett.} {\bf A278} (2001) 239.

\bibitem{ANANDAN}
J. Anandan, {\it Int. J. Theor. Phys.} {\bf 41} (2002) 199;
{\tt quant-ph/0304109}.

\bibitem{LIBSCHLICHENMAIER}
M. Schlichenmaier, {\it An Introduction to Riemann Surfaces, Algebraic Curves 
and Moduli Spaces}, Springer Lecture Notes in Physics {\bf 322}, Berlin (1989).

\bibitem{ARNOLD}
V. Arnold, {\it Mathematical Methods of Classical Mechanics}, Springer, 
Berlin (1989).

\bibitem{LIBAZCA}
J. de Azc\'arraga and J. Izquierdo, {\it Lie Groups, Lie Algebras, 
Cohomology and some Applications in Physics}, Cambridge University Press, 
Cambridge (1995).

\bibitem{CH}
R. Courant and D. Hilbert, {\it Methods of Mathematical Physics, I}, 
Interscience Publishers, New York (1953). 

\end{thebibliography}
\end{document}